\documentclass[runningheads]{llncs}
\usepackage[T1]{fontenc}
\usepackage{cite}
\usepackage{amsmath,amssymb,amsfonts}

\usepackage[noend]{algpseudocode}
\usepackage{tikz}
\usetikzlibrary{shapes.geometric}
\usetikzlibrary{positioning}

\usepackage{graphicx}
\usepackage{textcomp}
\usepackage{xcolor}
\usepackage{setspace}
\usepackage{placeins}
\usepackage{comment}
\usepackage{listings}
\usepackage{microtype}
\usepackage{hyperref}
\def\BibTeX{{\rm B\kern-.05em{\sc i\kern-.025em b}\kern-.08em
    T\kern-.1667em\lower.7ex\hbox{E}\kern-.125emX}}

\begin{document}

\title{Protecting Futures against Silent Data Corruption - Efficient Task Replication for Dynamic Data Dependencies
}

\titlerunning{Protecting Futures against SDC}

\author{Rüdiger Nather \and
Claudia Fohry \and
Mia Reitz}

\authorrunning{R. Nather et al.}
\institute{\textit{Research Group Programming Languages / Methodologies}\\ \textit{University of Kassel}\\ Kassel, Germany \\
\email{\{ r.nather~\textbar~fohry~\textbar~mia.reitz \}@uni-kassel.de}
}

\definecolor{comment-green}{rgb}{0.0, 0.5, 0.0}
\definecolor{keyword-blue}{rgb}{0.0, 0.0, 1.0}

\lstdefinelanguage{MyCustomLang}{
    morecomment=[l]{//},
    morekeywords={if, else},
    sensitive=true,
    showstringspaces=false,
    showspaces=false
}

\lstset{
    basicstyle=\small\ttfamily,
    commentstyle=\color{comment-green}\itshape,
    keywordstyle=\color{keyword-blue}\bfseries,
    numbers=left,
    xleftmargin=2em
} 

\maketitle

\begin{abstract}
    As the size of computational problems grows, so does the likelihood
  of Silent Data Corruptions (SDCs).  A common defense is replication,
  where the computation is repeated and correct results are determined
  by majority voting.  Asynchronous Many-Task (AMT) runtimes are generally well
  suited for this approach, since the inputs and outputs of task replicas can
  be compared, and the tasks can be recomputed if
  necessary. Most existing SDC protection schemes assume
  static tasks and dependencies. Dynamic settings are more
  challenging, especially in clusters, since the tasks/data must be tracked
  for the comparisons.

  This paper considers a particularly dynamic setting with task
  spawning at runtime, task communication through C++11-like promises/futures,
  conditional touches, and cluster-wide load balancing via
  work-first work stealing. We propose an approach that closely
  couples original and replica computations by cross-validating
  all outgoing effects when interacting with the runtime system. The
  approach selectively recomputes affected tasks only.

  We implemented the approach in the ItoyoriFBC runtime system and
  conducted preliminary experiments with Fibonacci and emulated \mbox{$\mathcal{H}$-matrix} LU decomposition
  benchmarks. Results show a factor of less than two increase of
  failure-free running times, despite full replication, which is
  mainly due to improved opportunities for load balancing resulting
  from the higher number of tasks. The overhead for failure correction
  was about 0.5\,\% of the overall running time per SDC.

\end{abstract}
\section{Introduction}\label{sec:introduction}

As HPC workloads grow in size and execution time, the likelihood of
Silent Data Corruptions (SDCs) increases.  SDCs are wrong results that
are caused by faults such as random bit-flips. They alter the program
state without triggering immediate failures, posing a critical threat
to result correctness.  A common defense is replication, in which the
computation is executed at least twice to detect failures. The
failures are afterwards corrected with the help of a third (full or
partial) execution, using majority voting.

Replication is well suited for Asynchronous Many-Task (AMT)
programming. In AMT, programs specify a large number of fine-grained
tasks, and the runtime system (RTS) maps the tasks to a smaller number
of execution units called workers. In clusters, the workers typically
correspond to one or several processes per node. Example AMTs include
HPX~\cite{Kaiser2014hpx}, ParSEC~\cite{parsec}, Legion~\cite{legion},
and ItoyoriFBC~\cite{itoFBC}.

The natural way of applying replication to AMTs is to compute
individual tasks twice, compare their results, and re-execute the tasks
if necessary. This has already been implemented for AMTs with static
tasks (e.g.~\cite{wang,samfass,ftx2,sriRajMPI}),
where comparisons between the results are straightforward. Other AMTs such as
HPX, Itoyori~\cite{itoyori}, and ItoyoriFBC~\cite{itoFBC}, however,
have \emph{dynamic tasks}, which are spawned at runtime, possibly
inside branches.  Dynamic tasks need to be
tracked to enable the comparisons. There is some recent work on SDC protection for
dynamic tasks~\cite{wamta26}. This work postpones failure
detection and repair to a postprocessing phase, at the price of
tripling the computational expense for early
failures. Moreover, the work in~\cite{wamta26} requires
all tasks to return their results to the parent, whereas AMTs often
permit directed acyclic
graph (DAG)-structured task dependencies.  Such dependencies were considered by Kurt
et al.~\cite{kurt}, but these authors presuppose direct dependencies
between tasks, whereas most AMTs express the dependencies indirectly
through data accesses.  Reference~\cite{kurt} also re-executes a
whole subgraph of the DAG instead of the infected task only.

This paper proposes an SDC protection scheme for dynamic tasks with
DAG-structured dependencies that are expressed through data accesses,
using the future construct. A future is a placeholder for a still to
be computed value.  All dependencies must be expressed with futures,
there are no side effects. We consider a recent, particularly flexible
variant of future~\cite{Nather2024futures,wamta25jour}, as well as a dynamic task
scheduling method, namely work-first work
stealing~\cite{cilkSchedule}, and a cluster environment.

The basic idea of our scheme is quite simple: We run the computation
twice and closely monitor the executions by cross-validating all
operations that cause effects outside their tasks.  These operations
(chiefly task spawns and writes to promises) involve the RTS, which records
every first occurrence of such an operation and releases it only after
having been confirmed by the other replica. Whenever the RTS detects a
mismatch, it creates a \emph{correction replica}, which repeats the
task computation from the beginning, but omits already executed
relevant operations such as spawns. When it arrives at the problematic
operation, it determines the correct version, unblocks the
corresponding task replica, and cancels the other.

Our scheme can detect and repair all processing-related SDCs inside
the application, except in the very unlikely case that two SDCs cause
exactly the same failure in two task replicas. SDCs in memory, in the
RTS, and during communication must be handled by other means, as in,
e.g.,~\cite{ftx1}. Furthermore, we assume that tasks are deterministic
regarding their outgoing effects.

We implemented the scheme in the ItoyoriFBC runtime
system~\cite{itoFBC, abstractMia} and performed preliminary experiments
with two benchmarks, Fibonacci and emulated $\mathcal{H}$-matrix LU decomposition, on up to
16~nodes with 640~workers. Interestingly, the overall running time increased by a factor of only
about~1.8 for \verb|fib|, and slightly below~2 for \verb|hlu|, despite
full duplication of the computation. We attribute this result to 
improved opportunities for load balancing due to a higher number of tasks and a reasonable degree of locality.  Moreover, we observed a correction overhead
per SDC of about~0.5\,\% of the overall running time on average, and
never above 2\,\%.
Altogether, our contributions are as follows:
\begin{itemize}
\item We propose a new SDC protection scheme for dynamic AMT
  runtimes under work-first work stealing in clusters. It closely couples two identical executions, timely
  detects mismatches, and selectively re-executes only the affected tasks.
\item We implemented a simple refinement of the scheme in the ItoyoriFBC cluster RTS and discuss possible improvements.
\item We conducted preliminary performance measurements without and with failures for two benchmarks. In these experiments, we
  observed low overheads.
\end{itemize}

The paper continues with background on AMT programming, future-based
task cooperation, and the ItoyoriFBC runtime system in
Section~\ref{sec:background}. Then it explains our new SDC protection
scheme in Section~\ref{sec:algorithm}. Thereby we first describe its
general idea, and afterwards sketch our implementation and discuss
alternatives. Thereafter, Section~\ref{sec:experiments} is devoted to
experiments, Section~\ref{sec:relWork} discusses related work, and
Section~\ref{sec:conclusions} concludes the paper.

\FloatBarrier
\section{Background}\label{sec:background}

\subsection{AMT Programming and Future Construct}\label{sec:progModel}

\begin{lstlisting}[label={lst:nfjfib},language=MyCustomLang,
  float=tp,frame=none,xleftmargin=2em,basicstyle=\small\ttfamily,numbers=left,caption=Naive Fibonacci computation in the FBC model]
void fib(int n, promise p0) {
  if (n < 2) {
    p0.set(1);
  } else {
    promise p1 = make_promise();
    spawn fib(n-1, p1);
    int a = fib(n-2);
    future f1 = p1.get_future();
    int b = f1.touch();
    p0.set(a+b);
  }
}
\end{lstlisting}

AMT programs specify a large
number of asynchronous tasks, and an RTS maps the tasks to a
smaller number of workers. This paper considers cluster
AMTs, where the workers correspond to one or several
processes per node.

Existing AMT environments differ in their task models, i.e., in their
mechanisms for task creation and
cooperation~\cite{fahringerSurvey,ichTut}. Common models include
Sequential Task Flow (STF) and Nested Fork-Join (NFJ). STF imposes a sequential order on task creation, and NFJ
requires child tasks to return their result to the parent.  Since
these models are quite restrictive, AMT environments often incorporate side effects. Unfortunately, side
effects undermine the clarity of task interfaces, which is an
important advantage of AMT and essential for failure protection.

This paper considers a lesser-known, but similarly expressive
programming model, called \emph{Future-Based Coordination}
(FBC~\cite{itoFBC}). It disallows side effects, but achieves
expressiveness with dynamic task spawning and the future construct.
An
FBC computation begins with a single root task.
Then, each task may spawn any number of children, recursively and possibly inside
branches.

Futures are single-assignment variables that can be communicated
before having been filled. They can be used for task synchronization,
whereby one task writes to a future, and others read from it, but are
blocked if necessary until the value is available.  This paper
considers a recent, particularly flexible variant of
future~\cite{Nather2024futures,wamta25jour}. Like C++11 or HPX
futures~\cite{c++Futs}, it is based on a placeholder memory location that can
separately be referred to for writing or reading. For writing, it is
denoted as \emph{promise}, and for reading as the actual
\emph{future}. Promises and futures can be stored in program variables
of types \verb|promise<T>| and \verb|future<T>|,
respectively. Thereby, \verb|T| may be any type, including an
incompletely defined one, which enables futures of futures.

After a promise has been written to, the promise and all
associated futures are \emph{ready}.  A future
is read by \emph{touch}ing it. If it is ready, the
touch returns the value. Otherwise, it suspends the calling task and the
executing worker switches to another task.

Both promises and futures may be passed as task parameters. Thereby
promises are assigned, meaning that the child task becomes responsible
for filling the promise. Futures, in contrast, are copied, such that
the child can read the future afterwards and pass it on to other
tasks.
The considered future variant enables the expression of highly dynamic dependencies due to several features:
\begin{itemize}
\item A task may start before its futures are ready. This allows it to decide internally within conditionals, which futures it actually needs to touch.
\item A future may be communicated to arbitrary, non-related tasks via parameter passing or by writing it into promises. Thereby the sender need not know the number and identities of the receivers.
\item Promises may be defined before task creation. Thus, the associated futures can be generated and communicated before it is clear which task will fill the promise.
\item A task that is responsible for filling a promise may decide
  during its execution to delegate this responsibility to another task.
\end{itemize}

Listing~\ref{lst:nfjfib} depicts FBC code for a naive Fibonacci
computation.  It is invoked by calling
\verb|fib(n, make_promise())|. The main branch starts with creating a
promise (\verb|make_promise|, line~5). Then it \verb|spawn|s a task
and assigns the responsibility for filling the promise by passing it
as a parameter (line~6). The associated future must be obtained
(\verb|get_future|, line~8), since only futures may be touched and
forwarded to other readers. This simple example does not show the
forwarding. Instead, the future is \verb|touch|ed by the parent task
(line~9). The touch call may suspend the task until the child task has
filled the promise with function \verb|set| (line~3 or~10). Afterwards
the sum is computed and written to the promise.

The code illustrates all relevant programming constructs, except for
the passing of futures, which is syntactically obvious. Note that the
passing of futures differs from that of promises, as
has been explained above. Reference~\cite{wamta25jour} lists a few
other constructs, but they can be emulated with the given ones.

\subsection{Work-First Work Stealing and ItoyoriFBC}

Dynamic tasks are often implemented with \emph{work stealing}. Therein
each worker maintains a local \emph{task pool}, for which we assume a
deque with a locking mechanism for access protection. Initially, all
pools are empty and one worker processes the start task. During normal operation,
a worker repeatedly takes the topmost task out of its pool and
processes it.  When the pool is empty (such as at the beginning),
the worker tries to \emph{steal} a task from a randomly chosen co-worker.
Thereby the \emph{thief} always takes the oldest
(lowermost) task from the \emph{victim} pool.

The behavior of a worker at a spawn is defined by the \emph{help-first}
or \emph{work-first} policy~\cite{workHelp}.  We consider
work-first, which has a strong theoretical foundation including
performance guarantees~\cite{cilkSchedule}. Under work-first, the
worker suspends the current task, constructs a \emph{continuation},
pushes the continuation onto its task pool, and branches into the child task.

Work-first work stealing has mostly been implemented on shared-memory
machines (e.g.~\cite{cilk5}), but a few cluster implementations exist
(e.g.~\cite{itoyori,kestor}). Notably, Itoyori~\cite{itoyori} and its
variant ItoyoriFBC~\cite{itoFBC} use a uni-address
scheme~\cite{uniAddress} to efficiently store and migrate
continuations.

These systems are C++ libraries and follow the NFJ or FBC model,
respectively. The newest version of ItoyoriFBC~\cite{abstractMia}
implements the future variant described above. It uses one-sided MPI
communication to realize a Partitioned Global Address Space (PGAS)
memory, which is not exposed to programmers, but only used in the
RTS. Among others, promises/futures are
stored in this memory, such that their addresses are global.  More
precisely, each promise is allocated in the local memory of the worker
that creates it and is never moved. When a task is suspended due to a
touch to a non-ready future, its continuation is inserted into a local
queue at the promise site. When the promise becomes ready, all
continuations from this queue are pushed onto the task pool of the
filling worker.

\FloatBarrier
\section{New SDC Protection Scheme}\label{sec:algorithm}

\subsection{General Scheme}\label{sec:general}

Summarizing Section~\ref{sec:progModel}, our FBC tasks can perform the following operations:
\begin{itemize}
\item standard sequential operations,
\item task spawns that include parameter passing of values, promises, and futures,
\item creations of promises and inquiries of associated futures,
\item writes to promises (briefly called psets), and
\item touches of futures.
\end{itemize}

Of these operations, only task spawns and psets cause data to leave
the boundaries of their task. Thus, our new SDC protection scheme
pursues the idea of confining failures to the task in which they
originally occurred by cross-validating these operations. More
specifically, the RTS delays their execution until they are identically requested by
two task replicas.

At program start, the scheme spawns two copies of the start task. One
of them is assigned to a worker for execution, and the other is placed
in the worker's task pool.  Two copies of the same task are called
\emph{twins} or \emph{replicas}.  The execution of twins is
coordinated with a \emph{task descriptor}, which contains:
\begin{itemize}
\item a pointer to the task code,
\item all task parameters,
\item in slots~1 and~2, the respective continuations and most recent
  unconfirmed operations of at most two suspended task replicas,
\item a counter for the number of previously committed operations, and
\item a lock to protect accesses to the task descriptor.
\end{itemize}

Task processing and work stealing are performed as usual, except for
spawns and psets.  When a task calls one of these functions, the RTS
executes the code in Listing~\ref{lst:recCommProt}, which
distinguishes different cases:

\begin{lstlisting}[label={lst:recCommProt},
    float=tp,xleftmargin=2em,basicstyle=\small\ttfamily,language=MyCustomLang,
  frame=none, numbers=left,caption={Outline of spawn/pset RTS functions (td = task descriptor, tp = task pool, R = replica, op = operation (spawn or pset), corr = correction, cont = continuation)}]
spawn(args) or pset(promise, value) {
  lock td;  
  if (both slots of td are empty) {
    record current cont and requested op in slot 1;
    assign top task from tp to worker (or start stealing);
  } else if (slot 1 of td is filled && slot 2 is empty) {
    if (equal(op in td, requested op)) {
      commit op (see below) and increase counter;
      if (op==spawn) assign child replica to worker;
      if (op==pset) return;
    } else {  // SDC has occurred
      record current cont and requested op in slot 2 of td;
      create corr-R and assign it to worker;
    }
  } else {
    // both slots of td are filled && R requested op
    if (counter in corr-R != counter in td) {
        increase counter in corr-R and return;
    } else if (equal(op in slot 1, requested op)) {
      commit op (see below);
      turn corr-R into normal R and assign it to worker;
    } else if (equal(op in slot 2, requested op)) {
      // analogous
    } else {
      abort;
    }
  }
  unlock td;
}

commit spawn(args) {
  create two child R's;
  push to tp: cont from td, current cont, one child R;
  clear both slots of td;
}

commit pset(promise, value) { // implementation-specific
  promise = value;
  resume conts of associated futures (as usual);
  move saved cont from td to tp;
  clear both slots of td;
}
\end{lstlisting}
 
First, when the descriptor does not currently contain a task replica,
the calling replica is ahead of its twin (lines~3--5). The RTS suspends
our task and \emph{records} its continuation and the requested
operation in the task descriptor. Then the worker continues with the
next task as usual, which typically is the other twin.

Second, when the descriptor already contains a suspended task (lines~6
to~14), its most recent operation (including parameters) is compared
with the currently requested one. Thereby the name of the operation,
its number of parameters, and all directly passed values must
agree.
This means that all relevant data must be held in memory in two locations,
until the correctness is confirmed by the task replica.
Furthermore, each passed promise and future must have the same type and contain the same address as its
counterpart. For the address, equivalence
may be sufficient for some implementations (see
Section~\ref{sec:discussion}).  In cases of full agreement, the
operation is \emph{committed}.

Committing of spawns is detailed in lines~31--35.  Notably
\emph{two} replicas of the child task are generated at once. Thus there are four
open tasks afterwards, of which three are inserted into the worker's
local pool (in this order):
\begin{itemize}
\item recorded continuation of twin task from task descriptor (bottommost),
\item continuation of current task,
\item one child replica (topmost).
\end{itemize}
Both continuations begin at the operation \emph{after} the spawn.  The
worker itself continues with the other child replica (line~9), according to the
work-first policy.

Committing psets may have to take the existence of two instances of
the promise into account, but our implementation uses a single instance
(see Section~\ref{sec:impl}). Therefore the
commit actions are analogous, but simpler than those for spawn
(lines~37--42).

 When the currently requested operation differs from the unconfirmed
 one in the task descriptor, an SDC must have occurred
 (lines~11--14). In this case, our replica is suspended and recorded in slot~2 of
 the task descriptor. Then, a \emph{correction replica} is
 created and assigned to the current worker. The correction
 replica repeats the task computation \emph{from the
 beginning}, using code and parameters from the task descriptor.

 Correction replicas are marked as such. During their execution, they \emph{skip} all spawns and
 psets until reaching the problematic operation (lines 17--18). Skipping is fine,
 since these operations have already been executed.  This way our
 scheme avoids the re-execution of healthy child
 tasks. For the problematic operation, the correction replica determines the
 correct one of the previously recorded versions, releases the
 corresponding continuation, removes the other, and commits the
 operation (lines~19--23).

 If the correction task disagrees with both previously recorded
 operations (lines~24--26), more than one SDC must have
 occurred within the same task, which is very unlikely. While a repair
 would be conceivable, we simply abort the overall program execution in this case.

\subsection{Implementation}\label{sec:impl}

For the implementation, we reconsidered the previous observation that
only spawns and psets cause data to leave the boundaries of their
task. While it is true, promise creations and touches also have outgoing
effects. Most importantly, a touch of a non-ready
future inserts the current continuation into a queue at the
promise. If it is the wrong promise due to an SDC, the continuation will not be
found by its twin and starve.

Our implementation resolves this issue in a simple way: It
records/commits not only spawns and psets, but also promise creations
and touches of futures. Like in Section~\ref{sec:general}, the first
replica that requests one of these operations records it and is
suspended, and the second replica commits the operation. 
Continuations are formed slightly differently here such that, after resuming,
they first read the address of the promise, or the value of the
future, respectively.

When a promise creation is committed, it allocates a \emph{single} placeholder
location. When a touch is committed, it returns the value if
available, or, otherwise, inserts the continuations of both replicas
into a queue at the promise.  Comparisons of promise creations check
the type of the promise, and comparisons of touches check type and
address of the future.

The placement of promises/futures in the cluster is the same as in ItoyoriFBC and, like
there, they are never moved.  Similarly, the task descriptor is
allocated in the local memory of the parent task and never moved.

\subsection{Discussion}\label{sec:discussion}

Our implementation is simple, but has drawbacks regarding
efficiency. Below we discuss its pros and cons and outline
optimization potential. Part of the discussion refers to the general
SDC protection scheme.

Local placement of the task descriptor at spawns tends to work well, since the
spawning worker directly branches into one child replica and often continues
with the other one afterwards (because it finds that replica topmost in
its task pool). Thus, as long as the child replica is not stolen, all accesses
to the task descriptor are local.

Steals of child replicas are fortunately rare, because these replicas are located opposite the
steal end in the task pool.  After child steals, however, the replica
must remotely access the task descriptor. Even worse, it is typically
behind its twin and thus commits the next operation. Thereby it moves
the saved continuation from the task descriptor to its own pool,
extending the remote accesses to the other replica.  A future
implementation may migrate the task descriptor at steals,
but we did not yet implement this optimization.

The locality is also limited:
Even though futures may be communicated to remote tasks,
the placement of promises is rigid. When a remote task
touches a non-ready future, its continuation is moved to the
promise. While this improves the locality of the touch itself,
accesses to the task descriptor become remote afterwards.  An improved
implementation should take the migration of promises and the
replication of filled promises into account.

The recording of promise creations and touches in our
implementation is simple, but introduces additional
suspension points. An alternative refinement of the general 
scheme could
create separate promises
in the twin computations and connect them to each other.
Separate promises have the additional
advantage that they enable the detection of
in-memory SDCs, which could afterwards be repaired
with the method from~\cite{wamta26}. As a second
alternative, promise creations and touches could be delayed until the
next spawn/pset and then be checked together with it. Lastly,
duplicate promises/futures could be removed at the next spawn/pset and
a conversion table for their addresses be maintained.

The SDC protection scheme itself differs from its non-resilient
counterpart by spawning two child replicas instead of one, which
necessarily increases the number of \emph{deviations} from the
standard sequential execution order. While deviations always reduce the
locality~\cite{acarLocality}, the particular deviations cause
relatively little harm because the 
corresponding code sections tend to be executed shortly after each other by
the same worker.

One may nevertheless question the frequent switching between tasks in our protection scheme. An alternative
scheme could duplicate only the code sections that correspond to these tasks and cross-validate their
operations inside the task. This may be preferable, especially for
code sections with a high density of relevant operations, but requires compiler
support. Moreover, the replication of tasks instead of code sections 
doubles the number of tasks.  In consequence, idle workers tend to find work
more quickly, which is particularly relevant in the starting phase of
a program execution.

This positive impact on load balancing can also be captured with the
well-known terms \emph{work} and \emph{span}. Using these terms, our
SDC scheme doubles the work, but it does \emph{not} double the span
because twin tasks may be run in parallel.  In fact, the spawn is not
increased at all at the program level, and it is only slightly
increased if one takes the additional RTS operations such as
comparisons into account.

Finally, correction replicas re-execute the affected
task itself, but not its children. Thus, the recovery overhead of our
scheme is at most $k$ task executions for $k$ SDCs.

\FloatBarrier
\section{Experiments}\label{sec:experiments}

\subsection{Experimental Setup}
\textbf{Hardware.}
All experiments were performed on a partition of the Goethe cluster of the University of Frankfurt~\cite{ClusterGoethe}.
This cluster consists of homogeneous Infiniband-connected nodes, each equipped with two 20-core Intel Xeon Skylake Gold 6148 CPUs and 192~GB of main memory.
We ran our experiments on up to~16 nodes with a total of up to 640~workers.
The source code of our experiments is available online~\cite{zenodo}.

\textbf{Software.}
We compiled our programs with Open~MPI~5.0.5 and g++~11.4.1 using the \texttt{-O3} switch.

\textbf{Benchmarks.}
We evaluated our scheme using two benchmarks with distinct characteristics:
\begin{itemize}
    \item \textbf{Naive Recursive Fibonacci (\texttt{fib}):} This benchmark resembles Listing~\ref{lst:nfjfib}, but spawns two child tasks instead of one. We use $n=62$ and a sequential cutoff of $C=32$, where calls to \texttt{fib}($n$) with $n<C$ do not spawn new tasks.
    \item \textbf{Hierarchical LU Decomposition (\texttt{hlu}):} This
      benchmark from~\cite{Nather2024futures} emulates the LU
      decomposition of hierarchical matrices. It spawns a quad-tree of
      tasks of height~$h$.  The emulation does not perform actual
      numerical computations, but the tasks busy-wait instead.  We set
      the busy-wait time to a total of~$t=100\,\text{s}$ across all
      tasks and use~$h=7$.
\end{itemize}

\textbf{Fault Injection.}
For \texttt{fib}, we simulated faults by injecting random bit flips into the task results set in promises.
For \texttt{hlu}, we simulated faults by causing the runtime to skip future touches.
Thereby, we test correctness in case of corrupted task results and control flow.

\textbf{Metrics.} Below \emph{Protection Overhead} denotes the percentage increase in running time of a failure-free protected run ($0$~SDCs) compared to a baseline unprotected run. \emph{Restore Overhead} denotes the percentage increase in running time of a protected run with one or more SDCs compared to a failure-free protected run.

Each experimental configuration was repeated 10 times, and we report the averages of these runs.

\subsection{Results}
\begin{figure}[t]
  \centering
  \input{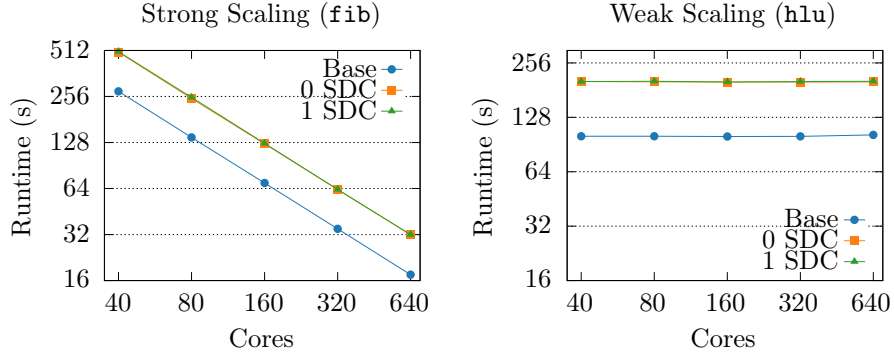}
  \caption{Strong scaling for \texttt{fib} (left) and weak scaling for \texttt{hlu} (right) in failure-free runs.}
  \label{fig:scaling}
\end{figure}

\begin{figure}[t]
  \centering
  \input{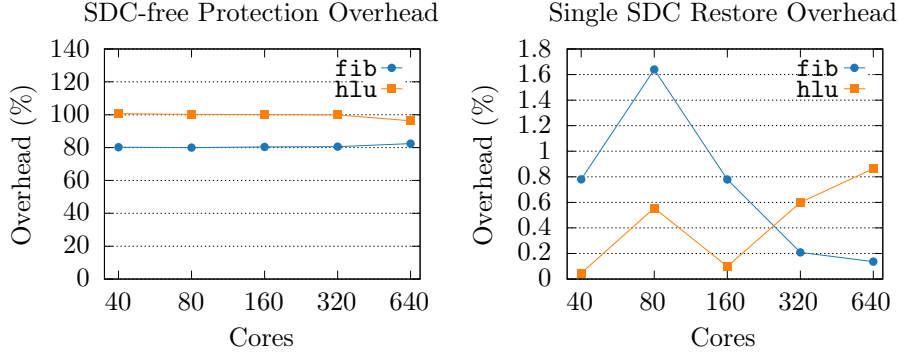}
  \caption{Protection overhead in failure-free runs (left) and restore overhead for a single SDC (right).}
  \label{fig:overhead-percent}
\end{figure}

\begin{figure}[t]
  \centering
  \input{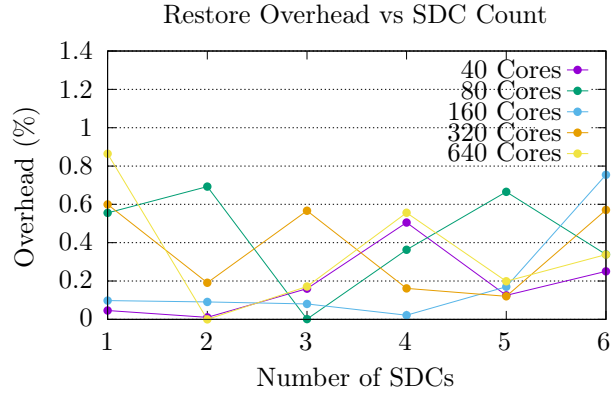}
  \caption{Restore overhead for multiple SDCs for \texttt{hlu}.}
  \label{fig:multi-sdc-hlu}
\end{figure}

Figure~\ref{fig:scaling} shows that the protected \texttt{fib} and
\texttt{hlu} programs have a similar scaling pattern to their
unprotected counterparts.  Figure~\ref{fig:overhead-percent} (left)
demonstrates that the failure-free protection overhead (0 SDCs) is
approximately $80\,\%$ for \texttt{fib} and slightly less than
$100\,\%$ for \texttt{hlu}. This outcome is remarkable, since one
would expect the protection overhead to be at least $100\,\%$, due to
the doubling of the task computations. The smaller overhead confirms
our argument from Section~\ref{sec:discussion} that the doubling of
tasks facilitates load balancing.

When faults are injected, the restoration cost is low.  As
shown in Figure~\ref{fig:overhead-percent} (right), restoring from a
single SDC adds less than $2\,\%$ overhead to the overall running time
for \texttt{fib} and less than $1\,\%$ for \texttt{hlu}.
Figure~\ref{fig:multi-sdc-hlu} illustrates that even when injecting up
to six SDCs during each run, the restore overhead is only about 0.5\,\%.

\section{Related Work}\label{sec:relWork}
AMT programming has received much attention
(e.g.~\cite{fahringerSurvey,ichTut}) and there is also some previous
work on fault tolerance. It partly addresses fail-stop errors and
handles them by techniques such as task-level
checkpointing and
supervision~(e.g.~\cite{posner}).

Some existing work on SDC protection relies on user-provided
plausibility tests or other application-specific methods for error
detection (e.g.~\cite{ftx1,ftx2}). Replication, in contrast, is a universal detection method.
It can be applied at different granularity levels, whereby
tasks are the natural level for AMTs~\cite{acr,wang}.

Replication can be used for both failure detection and repair. For repair,
all or part of the computation must be run three times and
the correct version be determined with majority voting
(e.g.~\cite{schneider,sriRajMPI}).  Depending on how quickly an error
is detected, re-running may be needed for the affected task only
(e.g.~\cite{ftx2}), or for an infected subgraph~\cite{kurt}. The computation can also be 
rolled back to a previous checkpoint~\cite{ftx1,subasiUnified}.

Thus far, replication has mostly been applied to static
tasks (e.g.~\cite{wang,samfass,ftx2,sriRajMPI}). For instance, Samfass et al. run
the task graph twice and reuse task results that have passed a
plausibility test. A recent proposal for dynamic tasks refers to NFJ
and postpones failure handling to a postprocessing
phase~\cite{wamta26}. Our approach, in contrast, detects and repairs
failures shortly after their occurrence to minimize wasted work. The
present paper was inspired by a similar, but still unimplemented
approach for help-first work stealing in shared-memory
environments~\cite{RUnpublished}.

\section{Conclusions}\label{sec:conclusions}

This paper has proposed a new SDC protection scheme for AMT programs with dynamic task spawns and
dynamic dependencies expressed with the future
construct. Its main idea is a close coupling of twin computations,
whereby operations with outgoing effects are recorded by the first
execution and committed after having been confirmed by the
second. In cases of mismatch, the affected task is repeated and the
correct operation determined by majority voting.

We first presented the scheme in a general way and then refined and
implemented it.  While the general scheme cross-validates only task
spawns and writes to promises, the refinement also cross-validates promise
creations and touches. Other refinements are conceivable and we
briefly discussed them. In preliminary experiments, we observed a performance
overhead of less than 100\,\% despite full duplication, due to
improved load balancing.

Future work may further improve this outcome by exploring refinements and investigating data redistribution.
Moreover, it should extend the experimental evaluation to more data-intense benchmarks.
Future work may also study other SDC protection schemes, which, for instance, perform comparisons at a coarser granularity, detect SDCs by checksums instead of replication, or map the replicas onto hardware to further increase the detection probability or to exploit mixed precision.

\FloatBarrier

\begin{credits}
\subsubsection{\ackname} 
This research was funded by the Deutsche Forschungsgemeinschaft (DFG, German
Research Foundation) under project number 512078735.
The authors gratefully acknowledge the computing time provided to them on the Goethe-NHR cluster at the Frankfurt Center for Scientific Computing.

\subsubsection{\discintname}
The authors have no competing interests to declare that are
relevant to the content of this article.
\end{credits}

\bibliographystyle{splncs03_unsrt}
\bibliography{lit}

\end{document}